\documentclass[journal]{IEEEtran}
\ifCLASSINFOpdf
\else
\fi

\usepackage{graphicx}
\usepackage{amsmath, amssymb}
\usepackage{algorithm}
\usepackage{algpseudocode}
\usepackage{booktabs}
\usepackage{tablefootnote}
\usepackage{cite}
\usepackage[caption=false,font=footnotesize]{subfig}

\makeatletter
\def\@IEEEdraftcls{1} % solo se draft
\makeatother
\IEEEaftertitletext{\vspace{-1em}}

\begin{document}
\title{A Game-Theoretic Decentralized Real-Time Control of Electric Vehicle Charging Stations - Part I: Incentive Design}

\author{Riccardo Ramaschi,
\IEEEmembership{Student Member, IEEE}, Mario Paolone, \IEEEmembership{Fellow, IEEE}, Sonia Leva, \IEEEmembership{Senior Member, IEEE}
\thanks{R. Ramaschi and S. Leva are with the Department of Energy, Politecnico di Milano, Milan, Italy. M. Paolone is with the Distributed Electrical Systems Laboratory, Swiss Federal Institute of Technology, Lausanne, Switzerland.}
}

\markboth{IEEE Transaction on Smart Grids}%
{Shell \MakeLowercase{\textit{et al.}}: Bare Demo of IEEEtran.cls for IEEE Journals}

\maketitle

\begin{abstract}
Large-scale Electric Vehicle (EV) Charging Station (CS) may be too large to be dispatched in real-time via a centralized approach. While a decentralized approach may be a viable solution, the lack of incentives could impair the alignment of EVs’ individual objectives with the controller’s optimum. In this work, we integrate a decentralized algorithm into a hierarchical three-layer Energy Management System (EMS), where it operates as the real-time control layer and incorporates an incentive design mechanism. A centralized approach is proposed for the dispatch plan definition and for the intra-day refinement, while a decentralized game-theoretic approach is proposed for the real time control. We employ a Stackelberg Game-based Alternating Direction Method of Multipliers (SG-ADMM) to simultaneously design an incentive mechanism while managing the EV control in a distributed manner, while framing the leadership-followership relation between the EVCS and the EVs as a non-cooperative game where the leader has commitment power. Part I of this two-part paper deals with the SG-ADMM approach description, literature review and integration in the abovementioned hierarchical EMS, focusing on the modifications needed for the proposed application. 
\end{abstract}


\section{Introduction}
Electric Vehicles (EV) sales are growing year by year, both globally - with a 35\% increase \cite{EVoutlook24} - and in Europe - with a 22.7\% share of new vehicle registration \cite{EEA} - in 2023. Concomitantly, private and public actors are investing on publicly accessible Charging Stations (CS) resulting, in Europe, in a 26\% increase for AC CS technology and a 53\% increase for DC CS technology during 2024 \cite{EAFO}.

In the context of Level 3 (L3) charging, Charging Point Operators (CPO) opt for CS with multiple Charging Points (CP) for three reasons, i.e. cost rationalization, competitiveness and profitability \cite{mckinsey}. Indeed, a well-established CPO for L3 charging like Tesla has an average number of CP per CS of 9.4 \cite{Tesla}. Another trend in L3 CS, emerging in light of the congestion issues linked with the deployment of Fast CS (FCS) \cite{Wang_impact}, is the integration with locally available Renewable Energy Sources (RES), e.g. Photovoltaic (PV), and stationary Battery Energy Storage Systems (BESS). Their integration on one hand complicate the CS operation but, on the other hand, pave the way for higher profitability and reduced grid congestions, provided that an Energy Management System (EMS) is in charge of the CS power exchange \cite{bess_pv}. In literature, EMS for PV\&BESS-powered FCS typically aims to reduce operational costs \cite{Kouka2020}, increase revenues \cite{RAMASCHI2024101531}, enhance infrastructure utilization \cite {Yao2017}, reduce grid congestion \cite{Grid_aware} or minimize emissions \cite{DR_Cabrera}. These objectives are often included in multi-objective optimization problems \cite{Mishra2023}, where user satisfaction and fairness frequently represent either additional objectives or metrics for the performance evaluation of the EMS \cite{Rudnik}. Indeed, the relationship between the involved agents, i.e. users (EVs) and CPO, is crucial to ensure efficient, fair, and reliable charging operations.

% \begin{table}[] \label{afir}
%     \caption{AFIR \cite{AFIR} Light-duty EV Public Charging Compliance Levels.}  
%     \centering
%     \begin{tabular}{l p{4cm}} % Adjust second column width
%         \toprule
%         \textbf{Requirement} & \textbf{Compliance Level} \\
%         \midrule
%         Power installed at end of year & 1.3 kW/BEV, 0.8 kW/PHEV \\
%         \midrule
%         TEN-T Core Network & 1 CS/60 km/direction \\
%         \textit{By Dec 31, 2025} & 400 kW, 1 point at 150 kW \\
%         \textit{By Dec 31, 2027} & 600 kW, 2 points at 150 kW \\
%         \midrule
%         TEN-T Comprehensive Network & 1 CS/60 km/direction \\
%         \textit{By Dec 31, 2027 (50\% Network)} & 300 kW, 1 point at 150 kW\\
%         \textit{By Dec 31, 2030} & 300 kW, 1 point at 150 kW \\
%         \textit{By Dec 31, 2035} & 600 kW, 2 points at 150 kW \\
%         \bottomrule
%     \end{tabular}
%     \label{tab:ev_requirements}
% \end{table}

\begin{table*}[t]
    \caption{Literature review on hierarchical ADMM applied in EV control}  
    \centering
    \begin{tabular}{c c c c} % Adjust second column width
        \toprule
        \textbf{Reference} & \textbf{Follower-Leader relation} & \textbf{Inter-layer communication} & \textbf{Agents} \\
        \midrule
        \cite{multi_ADMM} & Nested ADMM & Coupling constraints and aggregated power & Controller, aggregator and EVs\\
        \cite{similar_5} & Single-loop ADMM & Average profiles and dual variables & Aggregator and EVs (DSO included in constraints)\\
        \cite{similar_6} & Tri-layer exchange problem & Dual variables and aggregated power & DSO, aggregator and EVs\\
        \cite{similar_7} & Iterative ADMM & Dual variables and aggregated power & DSO, aggregator and EVs\\
        \cite{similar_concept} & Nested ADMM & Incentive-based signals as price and average profiles & Aggregator and EVs\\
        \cite{similar_2} & Nested ADMM & Incentive-based signals via adaptive penalties & CS and EVs\\
        \cite{similar_4} & Nested ADMM & Incentive-based signals via gradient projection & CS and EVs\\
        \cite{game_ADMM} & Non-cooperative game & Generalized Nash Equilibrium seeking algorithm & CS and EVs\\
        \cite{similar_3} & Non-cooperative game & Nash Equilibrium seeking algorithm & DSO and aggregators\\
        \bottomrule
    \end{tabular}
    \label{hierarchical_admm}
\end{table*}

These objectives can be tackled via centralized, decentralized or hierarchical approaches \cite{review_centr_decentr}. The choice of the approach is strongly dependent on the problem formulation, on the agents' relationships, on the scalability required and on the computational time requirements. For instance, in energy trading decentralized or hierarchical approaches are preferred \cite{Energy_trading}, while in FCS EMS the majority of literature uses a centralized approach to ensure a simpler control when a strong coordination is required. Nonetheless, those approaches raise concerns on computational time, privacy and information exchange. In \cite{Rudnik}, the authors proposed a real-time control for an EV CS that reduced the centralized problem complexity via a heuristic reducing the number of integer values. Another approach to reduce computational time, proposed in \cite{Grid_aware_2020}, is to distribute a separable centralized problem via the Alternating Direction Method of Multipliers (ADMM). This method is decentralized, offers low complexity and robustness, and it is particularly applicable in large-scale problems with reduced information exchange \cite{Boyd_ADMM}. Despite several works use this method for managing EV CS, i.e. \cite{Grid_aware_2020, ADMM_1}, the majority of the ADMM applications in this domain fall under the hierarchical approach \cite{Survey_ADMM}.

In this work, we develop a hierarchical three-layer EMS for an EV FCS, where the central controller is in charge of the day-ahead dispatch plan and the intraday refinement, while the real-time control relies on an ADMM-based method. The real-time control exploits on one hand the advantages of ADMM while, on the other hand, it takes into account the hierarchy among the central controller and the EVs. In literature, this relationship is often framed via multi-layer ADMM, as in \cite{multi_ADMM} where each agent and aggregator solve their specific distributed problem, or game-theoretic approaches, where the central controller is a Leader and the EVs are Followers \cite{game_ADMM}. In Table \ref{hierarchical_admm}, several articles using hierarchical-based ADMM are compared, focusing on the Follower-Leader relation modeling. The majority of the reviewed works adopt a multi-layer ADMM framework to structure the interaction among agents, wherein inter-layer communication either enforces coupling constraints (e.g., \cite{multi_ADMM, similar_5, similar_6, similar_7}) or conveys incentive-based signals (e.g., \cite{similar_concept, similar_2, similar_4}). On the other hand, \cite{game_ADMM} and \cite{similar_3} frames the relationship as a non-cooperative game, where the Leader responds to the Followers Individual Objectives (IO) trying to influence their behavior.

To the best of authors' knowledge, no study explicitly models the Followership and Leadership between the CS and the EVs as a Stackelberg Game (SG), a dynamic game where the Leader holds the commitment power by anticipating the follower reaction. This kind of modeling, proposed by Zheng et al. in \cite{CORE} in the context of big data network, introduce a SG-ADMM approach for the design of incentive mechanism via a two-layer formulation that will be described in Section \ref{sec2.1}. In Section \ref{sec2.2} the publications using this approach are critically reviewed, revealing that it exists a contextual research gap of the application of such approach in the real-time control of an EV CS. 

Our goal is to reduce the computational time required for the real-time operation while i) respecting the central controller setpoints' ranges, ii) ensuring optimality of the solution, and iii) designing a game-theoretic based incentive mechanism, balancing central and individual objectives. 
The novelty of this work consists in demonstrating the applicability - and the scalability - of such method in the context of electric mobility, where the SG-ADMM proposed in \cite{CORE} needs to be adapted not only to account for an incentive mechanism but also for the variable coupling constraint. In this first part of a two-part paper, we integrate SG-ADMM into the three-layer EMS in Section \ref{formul} and we propose some concluding remarks on the problem formulation in Section \ref{concl_part1}. In the second part of the paper, we will evaluate the proposed method in a specific case study.

\section{Stackelberg Game based Alternating Direction Method of Multipliers} \label{sec2}
\begin{figure*}[t]
    \centering
    \includegraphics[width=0.75\linewidth]{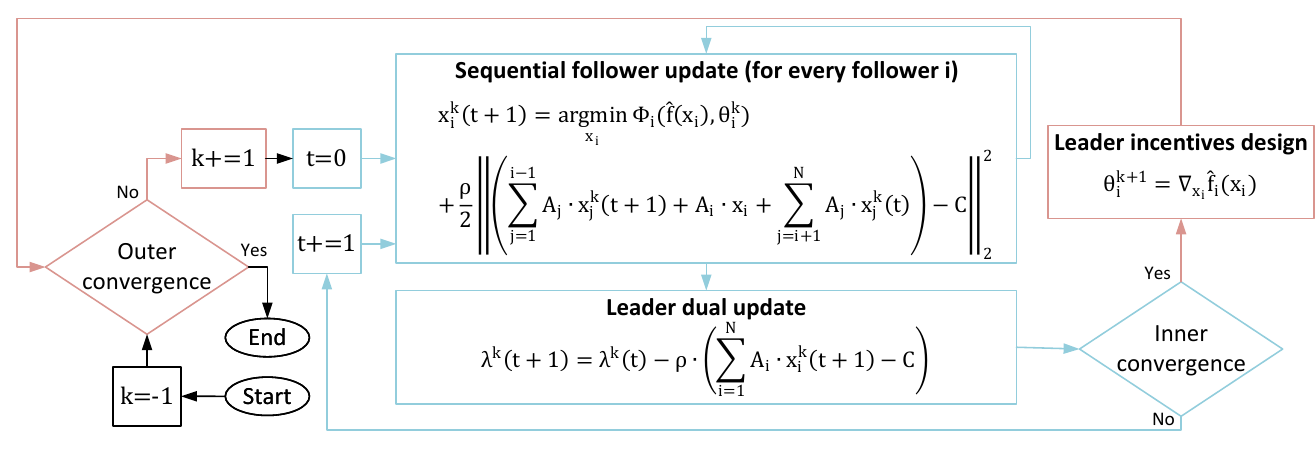} 
    \caption{SG-ADMM algorithm proposed in \cite{CORE} (in blue the ADMM inner loop, in red the SG outer loop).}
    \label{sg_admm_paper}
\end{figure*}
The proposed incentive mechanism design algorithm is meant to overcome the contrasting objectives between a central controller (a Leader) and the agents (the Followers). Indeed, if the followers do not update as the leader expects, a simple distributed optimization may not be able to reach the Leader optimum. To address this limitation, Zheng et al. \cite{CORE} introduced an incentive mechanism design framed as a Stackelberg game, where incentives are used to steer the Followers toward behaviors that align with the Leader’s objective.

\subsection{Incentive mechanism design algorithm} \label{sec2.1}
The Leader does not act on the decision variable $x_i$ but on the incentives $\theta_i$, that are meant to steer the Followers' optimum $\hat{x}_i$ towards the Leader's $x^*_i$. The algorithm will therefore be composed of a Leader Game, a Follower Game and a coupling constraint:
\begin{equation}
\begin{aligned}
&\hbox{Leader:} \quad \Theta^* =\operatornamewithlimits{argmin}_{\Theta} \sum _{i=1}^{N}f^*_i(x_i)\\ 
&\hbox{Followers:} \quad x_i^*=\operatornamewithlimits{argmin}_{x_i} \Phi_{i}(\hat{f}(x_i),\theta^*_i) \quad \forall \; 1\leq i \leq N, \\ 
&\hbox{Constraint:}\;\;\; \sum _{i=1}^{N} A_i \cdot x_i = C\\ 
\end{aligned} 
\end{equation}
Where $N$ is the number of the followers, $\Theta^*$ is the set of optimal incentives, $f^*_i$ and $\hat{f}_i$ are respectively the leader and the follower objective function, $\Phi_i$ is a Leader-designed incentive function, $A$ is a weight matrix and $C$ is the coupling constraint, linking the Leader to the Followers games.
The incentive function can be written as the sum of the follower-specific segmental Lagrangian function and a purely individual part:
\begin{equation} \label{incentive}
\begin{aligned}
    & \Phi_{i}(\hat{f}(x_i),\theta^k_i) = L_i(x_i,\lambda^k) + \phi^k_i(x_i)\\
     & L_i(P_i,\lambda) = f^*_i(x_i)-\lambda^k \cdot A_i \cdot x_i\\
     & \phi^k_i(x_i) = \hat{f}_i(x_i) - \theta^k_i \cdot x_i
     \end{aligned}
\end{equation}
Where $k$ represent the number of iteration for the Leader Game and $\lambda$ is the leader dual variable in the Follower Game. The design algorithm, comprising a two-layer iteration process, is shown in Figure \ref{sg_admm_paper}.

The convergence criteria are the following:
\begin{itemize}
    \item Inner loop:
    \begin{equation}
    \begin{aligned}
    & \|r^k(t+1)\|_2 \leq \epsilon'\quad \text{and} \quad \|s^k(t+1)\|_2 \leq \epsilon''\\
    &r^k(t+1) = \sum_{i=1}^{N} A_i\cdot x_i^k(t+1) - C\\
    &s^k(t+1) = \rho \sum_{i=1}^{N} A_i \cdot \left( x_i^k(t+1) - x_i^k(t) \right)
\end{aligned}
\end{equation}
Where $r^k$ and $s^k$ are the primal and dual residuals, respectively. $\epsilon'$ and $\epsilon''$ are the primal and dual residual tolerance. $\rho$ is the ADMM-specific penalty parameter.
    \item Outer loop:
    \begin{equation}
        \begin{aligned}
    & ||L(\textbf{x}^{k}, \lambda^{k}) - L(\textbf{x}^{k-1}, \lambda^{k-1})|| \leq \epsilon^L\\
    & L(\textbf{x}^{k}, \lambda^{k}) = \sum^{N}_{i=1} L_i(x^k_i,\lambda^k) + \lambda^k \cdot C\\ & \qquad \qquad  + \frac{\rho}{2} \left\| \sum_{i=1}^{N} A_i \cdot x_i  - C\right\|_2^2
        \end{aligned}
    \end{equation}
That is a primal stopping criterion for the leader Lagrangian, where $\epsilon^L$ is the corresponding tolerance.
\end{itemize}
\subsection{Literature review: SG-ADMM for single-leader multi-followers non-cooperative games} \label{sec2.2}
Since the method showcased in Section \ref{sec2.1} has been applied in different contexts and architectures, we analyze the literature on application of SG-ADMM for single-leader multi-followers non-cooperative games, as our problem. We found 13 papers using the same - or similar - SG-ADMM model explained above, falling in two different macro-areas.

\begin{figure*}[ht]
    \centering
    \includegraphics[width=0.97\linewidth]{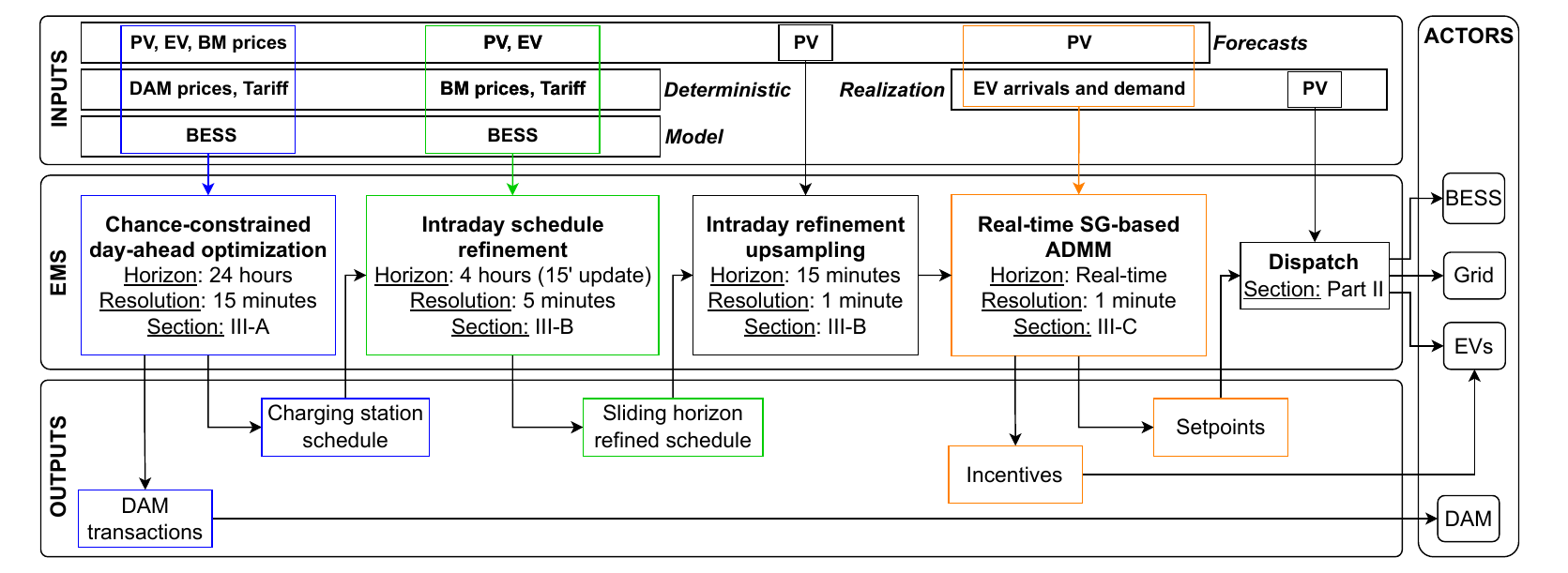} 
    \caption{Proposed framework for the optimal EMS of a CS. A three-layer EMS is proposed, where only the last layer, the Real-time SG-based ADMM is the focus of the paper.}
    \label{framework}
\end{figure*}

\subsubsection{Big Data and Edge Computing}
Zheng et al. firstly proposed this method in \cite{CORE, 1_1} to overcome the issue of contrasting objectives among data server (leader) and mobile devices (followers). Other publications by Zheng (as a first author) applied the same method to large scale mobile edge networks \cite{1_2} and edge caching \cite{1_3}, focusing on service providers need to develop a pricing mechanism for edge nodes storage and backhaul resources. Authors on \cite{1_4} focus on Intelligent Reflecting Surfaces (IRS) aided communications, using this method to solve the reluctance to help the base station of the IRS, that may belong to other operators, without any incentive. Similarly, \cite{1_5} addressed the same problem in the context of mobile edge computing for healthcare systems, with an additional aim of reducing the overall energy cost by means of the incentive mechanism. Also \cite{1_6} addressed mobile edge computing in the context of education the COVID-19 pandemic, to overcome huge bandwidth usage and unpredictable latency. Lastly, \cite{1_9} aims to design an incentive mechanism that steer participants to obtain optimal training and mining outcomes in a federated learning blockchain framework.

\subsubsection{Internet of Things and Smart Cities}
The uptake of Internet of Things (IoT) enabled a large amount of processing and storing capability. Nonetheless, effective incentives should be designed to convince the IoT devices to participate in cloud computing. Thus, several papers used the SG-ADMM concept for energy efficiency maximization and network latency minimization \cite{2_1} and optimal allocation for fog computing \cite{2_2}. A particular group of IoT devices, named Internet of Vehicles (IoV), have also been investigated from the point of view of computing offloading in parked vehicles \cite{1_8}, and aerial-assisted IoV \cite{2_3, 2_4}. Strictly linked to IoT is the concept of Smart Cities. In fact, \cite{3_1} analyzes the joint optimization of energy conservation and privacy preservation for intelligent task offloading in Smart Cities with a high penetration of mobile edge computing. Lastly, \cite{3_2} introduces the selfishness of EVs when it comes to where and how to charge. Here the smart city coordinator (leader) aims to design optimal price functions of CSs and a traffic coordinator to optimize the social welfare. The EVs (followers) optimally decide on their route and charging destination according to the price signal.

While most of the works deploy SG-ADMM in computer and communication engineering, this review highlighted its applicability in any context where a leader and multiple followers have contrasting objectives that need to be reconciled for achieving a satisfying result. This is the reason SG-ADMM is chosen in this work as the real-time control of an EV CS, so that the CS designs a incentive mechanism - holding the abovementioned commitment power - to align Followers IOs to the CS objectives. The result is a unified formulation of both objectives solved with a game-theoretic distributed algorithm.

\section{Problem Formulation} \label{formul}
Figure \ref{framework} shows the proposed framework for the overall EMS of the CS. Despite the focus of this paper is the real-time SG-based ADMM, that will be explained in section \ref{core}, we defined a comprehensive framework for the optimal operation of a CS, where the SG-ADMM plays a crucial role in the optimal management of the system. Briefly, a Chance-Constrained Day-Ahead model generates the dispatch plan over the whole day (see Section \ref{cc_da}), that is refined via an Intraday Schedule Refinement (see Section \ref{id_s_ref}). This refinement is upsampled so that the Real-time SG-ADMM control is provided of the needed constraints and inputs (see Section \ref{core}). Before introducing these three layers, the CS modeling is proposed in Section \ref{model}.

% \begin{figure}[b]
%     \centering
%     \includegraphics[width=0.1\linewidth]{Images/CS.png}
%     \caption{Charging station layout: components and configuration. The image is adapted from \cite{pt} with authors' permission.}
%     \label{charging_station}
% \end{figure}

\subsection{Charging station modelling} \label{model}
The CS, whose layout is presented in \cite{pt}, plays in the electricity market, bidding in the Day-Ahead Market (DAM) a Dispatch Plan (DP) and operating in real-time in the Balancing Market (BM). The CS purchase (or sell) energy in the DAM according to the market clearing rate ($\mathcal{R}_{\text{DAM}}$). In the BM, the deviation of the CS power exchange with the grid with respect to the DP are penalized (or remunerated) at a certain rate ($\mathcal{R}_{\text{BM}}$). For reducing the amount of energy withdrawn with respect to the DAM, a long rate is applied ($r_{long}$), while for increasing the amount of energy withdrawn a short rate is applied ($r_{short}$). The EV pricing scheme comprises an energy charge ($\mathcal{C}$) dependent on the period, $c_{p}$ in peak periods and $c_{op}$ in off-peak periods. On top of the energy charge, a potential discount is applied in real-time according to the designed incentive mechanism, as described in Section \ref{core}.

Considering typical power ratings of CSs, it is supposed to be connected to the Medium Voltage (MV) grid, and it is equipped with a PV field and $n_{CC}$ unidirectional Charging Columns - each of them with a $P_{CC}$ nominal power and $n_{CP}$ L3 CPs. Moreover, the CS is also supposed to be equipped with a stationary BESS, modeled as follows, using a standard lossy bucket representation:
\begin{equation}
\begin{aligned}
    & P^{\eta}_{B,i} = \frac{P^{ch}_{B,i}}{\eta_{inv} \eta_{ch}}\cdot \delta_i + P^{dh}_{B,i}\eta_{inv} \eta_{dh}\cdot (1-\delta_i)\\
    & = \begin{cases}
        \frac{P^{ch}_{B,i}}{\eta_{inv}\eta_{ch}} \geq 0 & \text{if } \delta_i = 1 \quad \text{(Charging)}\\
        P^{dh}_{B,i}\eta_{inv} \eta_{dh} \leq 0 & \text{if } \delta_i = 0 \quad \text{(Discharging)}
    \end{cases}\\
    & \Delta SoC_{B,i\%} = \frac{\left(P^{ch}_{B,i}\cdot \delta_i+P^{dh}_{B,i}\cdot (1-\delta_i)\right) \Delta i}{C_{B}}\cdot100\%.
\end{aligned}
\end{equation}
Where $i$ is a general time step, $P^{ch}_{B}$ and $P^{dh}_{B}$ are the charging and discharging power at the battery level, respectively, while $P^{\eta}_{B}$ is the corresponding power at the AC bus level. Since charging and discharging cannot happen simultaneously, $\delta$ is a binary variable indicating the power flow. $\eta_{inv}$, $\eta_{ch}$ and  $\eta_{dh}$ are the BESS inverter, and cells' charging and discharging efficiencies, respectively (all having a value $<1$). Lastly, $\Delta SoC_{B\%}$ is the percentage variation of the State of Charge (SoC), that depends on the exchanged energy (numerator) and the battery energy capacity ($C_{B}$). $\Delta i$ is the considered time interval.

The degradation model considered for the stationary and the EVs' batteries is taken from \cite{Shubham}. The empirical model utilizes a Stress Factor (SF) based approach to compute the degradation factor in a specific condition with respect to a reference degradation factor. In this work, only the cycling aging will be considered, with degradation computed as follows:
\begin{equation} \label{SF}
    \begin{aligned}
        & \qquad d^{act}_{B} = d^{ref}_{B} \cdot SF\\
        & \qquad SF = SF_{SoC} \cdot SF_{temp} \cdot SF_{DoD} \cdot SF_{Cr}\\
    \end{aligned}
\end{equation}

$d^{act}_{B}$ and $d^{ref}_{B}$ are the actual and the reference cycling degradation for the BESS. They are linked by the stress factor $SF$, product of the stress factors for each of the four variables ($SF_{SoC}$ for SoC, $SF_{temp}$ for temperature, $SF_{DoD}$ for Depth of Discharge (DoD), and $SF_{Cr}$ for C-rate), that can be found in \cite{Shubham}. Since $d^{act}_{B}$ is defined as a percentage degradation per Full Equivalent Cycle (FEC) ($\%/FEC$), the absolute degradation is computed as the product of the former for the number of FEC ($N_{FEC}$):
\begin{equation} \label{degradation}
    d_{B} = d^{act}_{B} \cdot N_{FEC}
\end{equation}

The CS is also supposed to be equipped with forecasters for PV production, EV demand and BM prices. These forecasters, feeding information to the three-layer EMS, are discussed in the second part of this paper.

\subsection{Chance-Constrained Day-Ahead optimization} \label{cc_da}
The first layer of our EMS is a chance-constrained DA model providing a dispatch plan to the DAM with a robust objective over $\mathcal{T}_{DA}$, the set of time periods where $T$ represents the timestamps and $\Delta T$ the interval. An imbalance price estimation is also performed to define an actual schedule for the power exchange. $\hat{r}_{long,T}$ and $\hat{r}_{short,T}$ are the estimation for the long and the short imbalance price in the balancing period $T$.

The objective of this optimization is to maximize the profit considering the charging revenues and the expected value of the cost of energy (considering both the DAM and the BM), both accounting for the stochastic variables effect. An additional term aims to reduce the BESS throughput considering its cycling degradation.
\begin{equation}
\begin{aligned}
\operatornamewithlimits{min}_{P_{EV,T},P^{\eta}_{B,T}} 
\mathop{\displaystyle \mathbb{E}}&
\bigg[ 
\sum_{T \in \mathcal{T}_{DA}} C_{DAM,T} 
- R_{EV,T}\\
&+ h(P^{\eta}_{B,T}) + C_{BM,T}
\bigg]
\end{aligned}
\end{equation}

Where $C_{DAM,T}$ and $C_{BM,T}$ are the costs associated with the DAM and BM, $R_{EV,T}$ are the revenues from EV charging and $h(P^{\eta}_{B,T})$ links the BESS power exchange in certain conditions to a certain degradation ($d_{B}$ from Eq. \ref{degradation}), that is in turn made an expense. The function is defined as follows:
\begin{equation} \label{BESSdeg}
\begin{aligned}
    & h(P^{\eta}_{B,T}) = \frac{d_{B}}{D_{B,EoL}} \cdot C_{B} \cdot p_{kWh}
\end{aligned}
\end{equation}
Where $D_{B,EoL}$ is the BESS degradation at End of Life (EoL), $p_{kWh}$ is the BESS energy price and variables $\bar{DoD}, \bar{SoC}, \bar{Cr}$ are the average value of DoD, SoC and C-rate - respectively - during one time step. The temperature, present in the stress factor in Eq. \ref{SF}, is considered constant as it is outside of our scope. 

$C_{DAM,T}$, $C_{BM,T}$ and $R_{EV,T}$ are formulated as follows:
\begin{equation} \label{obj_1st}
\begin{aligned}
    &  C_{DAM,T} = P^{dp}_{G,T} \cdot \mathcal{R}_{DAM,T}\\
    & C_{BM,T} = \hat{r}_{short, T}\cdot (P_{G,T}-P^{dp}_{G,T})^+\\ 
&\phantom{C_{BM,T}..}+ \hat{r}_{long, T}\cdot (P^{dp}_{G,T}-P_{G,T})^+\\
& R_{EV,T} = P_{EV,T} \cdot \mathcal{C}_T
\end{aligned}
\end{equation}

$P_{EV,T}$ is the cumulated EV satisfied demand, $P^{dp}_{G,T}$ is the dispatch plan submitted in the DAM, while $P_{G,T}$ is the CS internal power schedule. We define as $\Delta P^\pm$ the signed deviations between the two variables. When the long and short rate are positive, the dispatch plan is equal to the internal power schedule. The relationship between the two can be written down as:
\begin{equation}
    P_{G,T} = P^{dp}_{G,T} + \Delta P^+_T - \Delta P^-_T
\end{equation}

The model is constrained as follows:
\begin{align}
    &P_{G,T} \cdot \eta_{tr} + \bar{P}_{PV,T} \cdot \eta_{pv} = \frac{P_{EV,T}}{\eta_{cp}} + P^{\eta}_{B,T} \label{eq1} \\
    &|P_{G,T}| \leq P_{GC} \label{eq2} \\
    &|P_{B,T}| \leq C_B \cdot c_{rate} \label{eq3} \\
    & SoC_{min} \leq SoC_{B,T} \leq SoC_{max} \label{eq5} \\
    & P_{EV,T} \leq \bar{P}_{EV,d,T} \label{eq4}
\end{align}

Eq. \ref{eq1} represents the power balance of the AC CS main busbar, where $\bar{P}_{PV,t}$ is the stochastic variable for the PV realization. $\eta_{tr}$, $\eta_{cp}$ and $\eta_{pv}$ are the efficiencies for the grid transformer, CP converter (since we refer to a L3 CS) and PV converter, respectively. Eq. \ref{eq2} and \ref{eq3} threshold the grid and BESS power exchange to their operational bounds. In Eq. \ref{eq3}, despite the C-rate is SoC-dependent, we consider a conservative lower bound for C-rate ($c_{rate}$), taking the minimum C-rate in the operating ranges defined in Eq. \ref{eq5}.  Eq. \ref{eq4} limits the EV satisfied demand to the stochastic EV demand $\bar{P}_{EV,d,t}$.

Besides the BM prices, EV demand and PV production forecasts for this model are expressed as $\hat{P}_{EV,T}$ and $\hat{P}_{PV,T}$. We obtain the median value but also a certain range ($[^{\downarrow}; ^{\uparrow}]$) according to a predefined confidence interval of the forecaster. We chance-constrain the stochastic realization for the PV production and EV demand according to a predefined confidence interval. We separate the constraints according to \cite{boyd}, in order to ensure they are satisfied in any condition within the confidence level. We therefore obtain the following separable constraints for grid withdrawal, grid injection and EV demand:
\begin{align}
    &\frac{1}{\eta_{tr}}\left(\frac{P_{EV,T}}{\eta_{cp}} + P^{\eta}_{B,T} \right) \leq \hat{P}^{\downarrow}_{PV,T} \cdot \eta_{pv} + P_{GC}\label{withdraw}\\
    &\frac{1}{\eta_{tr}}\left( \frac{P_{EV,T}}{\eta_{cp}} + P^{\eta}_{B,T} \right) \geq \hat{P}^{\uparrow}_{PV,T} \cdot \eta_{pv} - P_{GC} \label{inject}\\
    & P_{EV,T} \leq \hat{P}^{\downarrow}_{EV,d,T} \label{evdem}
\end{align}

To hedge against excessive speculation in the DAM, a speculation factor $f_{s}$ has been introduced. This factor limits the amount of available power to bid on the DAM for selling (Eq. \ref{sell}) and buying (Eq. \ref{buy}) energy, in a risk-averse-like heuristic approach.
\begin{align}
    &P_{G,T}^{dp} \geq (SoC_{min} - SoC_{B,T}) \cdot f_s - \hat{P}_{PV,T} \cdot \eta_{pv} \label{sell}\\
    &P_{G,T}^{dp} \leq (SoC_{max} - SoC_{B,T}) \cdot f_s + \frac{\hat{P}_{EV,T}}{\eta_{cp}}\label{buy}
\end{align}
Following the model fine-tuning phase, the speculation factor was set to 80\%, intentionally limiting market exposure to maintain 20\% BESS flexibility for risk management.

We can now write the complete DA model as follows:
\begin{equation} \label{objf_final}
\begin{aligned}
&\operatornamewithlimits{min}_{P_{EV,T},P^{\eta}_{B,T}} \sum_{T \in \mathcal{T}_{DA}} P^{dp}_{G,T} \cdot \mathcal{R}_{DAM,T} - P_{EV,t} \cdot \mathcal{C}_T + h(P^{\eta}_{B,T})\\
&  + \hat{r}_{short, T}\cdot (\frac{P_{EV,T}}{\eta_{cp}} + P^{\eta}_{B,T} - \hat{P}_{PV,T} \cdot \eta_{pv}-P^{dp}_{G,T})^+\\ 
& + \hat{r}_{long, T}\cdot (P^{dp}_{G,T}-\frac{P_{EV,T}}{\eta_{cp}} + P^{\eta}_{B,T} - \hat{P}_{PV,T} \cdot \eta_{pv})^+\\
   &\text{s.t. (\ref{eq3}), (\ref{eq5}), (\ref{withdraw}), (\ref{inject}), (\ref{evdem}), (\ref{sell}), (\ref{buy})}
\end{aligned}
\end{equation}

\subsection{Intra-day schedule refinement} \label{id_s_ref}
As mentioned in the description of the three-layer EMS, an optimized charging station playing in the balancing market must optimize its energy exchange according to the imbalance prices and the updated available information. The goal of this procedure is to refine the day-ahead schedule according to the realization and the updated forecasts.

We adopt a sliding window approach to refine, at the beginning of each BM session (therefore with an update time equal to the DA problem granularity $\Delta T$), the CS schedule. We optimize over an horizon $\mathcal{T}_{BM}$ with a granularity $\Delta t$ (and $t$ as a time step). The goal of this procedure is to define for the refinement horizon a BESS power scheduling ($P^{\eta}_{B,t}$) and the grid power budget range ($P^{\downarrow}_{G,t}$, $P^{\uparrow}_{G,t}$). The concept of grid power budget is derived from the problem formulation in \cite{Power_budget}. The inputs are the maximum and expected EV power demand and PV forecast range, defined as follows:
\begin{enumerate}
    \item Maximum power demand, $P^{max}_{EV,d,t}$, is obtained via the booking system, having the same horizon $\mathcal{T}_{BM}$. Expected power demand, $\hat{P}_{EV,d,t}$, is obtained via an intraday forecaster.
    \item PV forecast range, $[\hat{P}^{\downarrow}_{PV,t}, \hat{P}^{\uparrow}_{PV,t}]$, is obtained via an intraday forecaster with a certain confidence interval. $\hat{P}^{\downarrow}_{PV,t}$ is the lower quantile, $\hat{P}^{\uparrow}_{PV,t}$ is the higher quantile and $\hat{P}_{PV,t}$ is the median.
\end{enumerate}

Morover, cost and tariff are still considered deterministic, as well as the short and long rate ($r_{short,t}$, $r_{long,t}$) over $\mathcal{T}_{BM}$.

We define the grid power budget range as follows, where $s^{-}_{t}$ and $s^{+}_{t}$ are the negative and positive EV slack variables and $\bar{P}_{G,t}$ is the expected grid power inside the budget range:
    \begin{equation}
    \begin{aligned}
        & P^{\downarrow}_{G,t} \cdot \eta_{tr} = \frac{P_{EV,t} - s^{-}_{t}}{\eta_{cp}} + P^{\eta}_{B,t}-\hat{P}^{\uparrow}_{PV,t} \cdot \eta_{pv}\\ & 
        P^{\uparrow}_{G,t} \cdot \eta_{tr} = \frac{P_{EV,t}+ s^{+}_{t}}{\eta_{cp}} + P^{\eta}_{B,t}-\hat{P}^{\downarrow}_{PV,t} \cdot \eta_{pv}\\
        & P^{\downarrow}_{G,t} \leq P_{G,t} \leq P^{\uparrow}_{G,t}\\
        & \bar{P}_{G,t} \cdot \eta_{tr} = \frac{P_{EV,t}}{\eta_{cp}} + P^{\eta}_{B,t}-\bar{P}_{PV,t} \cdot \eta_{pv}\\
    \end{aligned}
    \end{equation}

The optimization considers four objective, described as follows:
\begin{enumerate}
    \item Imbalance price minimization. The intraday refinement provide $\frac{\Delta T}{\Delta t}$ grid values inside each BM update, therefore the expected positive and negative deviation from the DP are generally defined as follows:
    \begin{equation}
            E_{G, T} = P^{dp}_{G,T} \Delta T + \bar{E}_{G,T}^+ - \bar{E}_{G,T}^- = \sum^{t+\frac{\Delta T}{\Delta t}}_{t} \bar{P}_{G,t} \cdot \Delta t
    \end{equation}
    Therefore the minimization of the imbalance price will be as follows:
    \begin{equation}
        \sum_{T \in \mathcal{T}_{BM}} \left(\bar{E}_{G,T}^+ \cdot r_{short,T} +  \bar{E}_{G,T}^- \cdot r_{long,T}\right)
    \end{equation}
    \item EV slack variable minimization. It aims at minimizing the total slack sum while also reducing their L2-norm to prevent imbalances.
    \begin{equation}
        \sum_{t \in \mathcal{T}_{BM}}  \left(s^{+}_t+s^{-}_{t} + \left(s^{+}_t-s^{-}_{t}\right)^2\right)
    \end{equation}
    \item SoC tracking. This term aims at tracking the predefined BESS schedule from the DA problem for the first and the last step inside $\mathcal{T}_{BM}$.Thus, this term can be written as:
    \begin{equation}
    \begin{aligned}
    f_{SoC,T} &= (SoC_{B,\mathcal{T}_{BM}} - SoC^{dp}_{B,\mathcal{T}_{BM}})^2 \\
            &\quad + (SoC_{B,T} - SoC^{dp}_{B,T})^2
    \end{aligned}
    \end{equation}
    \item Profit maximization.
    \begin{equation}
        \sum_{t \in \mathcal{T}_{BM}} P_{EV,t}\cdot \Delta t\cdot  \mathcal{C}
    \end{equation}
\end{enumerate}

The objective function can therefore be written as follows:
\begin{equation}
\begin{aligned}
\operatornamewithlimits{min}_{P_{t},s_{t}} & \sum_{T \in \mathcal{T}_{BM}} \left(\bar{E}_{G,T}^+ \cdot r_{short,T} +  \bar{E}_{G,T}^- \cdot r_{long,T}\right)\\
&+ b \cdot\sum_{t \in \mathcal{T}_{BM}} \left(s^{+}_t+s^{-}_{t} + \left(s^{+}_t-s^{-}_{t}\right)^2\right)
+ \sum_{t \in \mathcal{T}_{BM}} \Delta h(P^{\eta}_{B,t})\\
& + c \cdot f_{SoC,T} - \sum_{t \in \mathcal{T}_{BM}} P_{EV,t}\cdot \Delta t\cdot  \mathcal{C}
\end{aligned}
\end{equation}
Where $b$ is the weight for the slack term, $c$ is the penalization factor for the SoC tracking term. The minimization is performed according to the set of decision variables $P_{t}=\{P^{\eta}_{B,t}, P_{EV,t}\}$ and $s_{t}=\{s^+_{t},s^-_{t}\}$.

This optimization problem is constrained by general purpose constraints:
\begin{equation}
    \begin{aligned}
        & P^{\uparrow}_{G,t} \leq  P_{GC}\\
        & P^{\downarrow}_{G,t} \geq  -P_{GC}\\
        & \vert P^{\eta}_{B,t} \vert \leq C_{B}\cdot c_{rate}\\
        & SoC_{min} \leq SoC_{B,t} \leq SoC_{max}\\
    \end{aligned}
\end{equation}

Other constraints, instead, involve EV demand and its slacks:
    \begin{equation} \label{slacks}
        0 \leq P_{EV,t} - s^{-}_{t} \leq \hat{P}_{EV,d,t} \leq P_{EV,t} + s^{+}_{t} \leq P^{max}_{EV,t}\\ 
    \end{equation}
    \begin{equation} \label{tau}
        \begin{aligned}
        & \frac{s^{+}_{t} + s^{-}_{t}}{\eta_{cp}} + \Delta P_{PV,t}\cdot \eta_{pv} \geq \tau \cdot \left(\bar{P}_{G,t}\cdot\eta_{tr} - P^{\eta}_{B,t}\right) \\
        & \Delta \hat{P}_{PV,t} = \hat{P}^{\uparrow}_{PV,t} - \hat{P}^{\downarrow}_{PV,t}  \\
        \end{aligned}
    \end{equation}
Eq. \ref{slacks} sets $P_{EV,t}$ to be in the physical feasible space, its range lower bound $P_{EV,t} - s^{-}_{t}$ to be between 0 and $\hat{P}_{EV,d,t}$ and its range upper bound $P_{EV,t} + s^{+}_{t}$ to be between $\hat{P}_{EV,d,t}$ and $P^{max}_{EV,d,t}$. Eq. \ref{tau} imposes a lower bound on the width of the flexibility range provided by the EVs $ \frac{s^{+}_{t} + s^{-}_{t}}{\eta_{cp}}$ and the PV generation uncertainty $\Delta P_{PV,t}$. The constraint ensures that this available flexibility is sufficient to accommodate a proportion $\tau$ - defined by the decision maker - of the expected power made available to the EVs by the grid and the BESS. During the model fine-tuning, $b$ was settled to 1 and $c$ to 0.01. This choice, obtained after testing multiple combinations, reflect the importance given to each term. That is, $s^+_{t}$ and $s^-_{t}$ should be soft-constrained to their lower bound and as close as possible, while SoC tracking should be a weaker term allowing to deviate from the DP in case relevant economic opportunities occur in the BM. At the same time, to allow a significative flexibility band, $\tau$ is set to 20\%.

The sliding-window approach becomes a shrinking horizon approach when $\mathcal{T}_{BM}$ falls outside of the operation day. In that case, $\mathcal{T}_{BM}$ is set to the last DA time step $\mathcal{T}_{DA}$.

The outputs of this optimization are the BESS power scheduling ($P^{\eta}_{B,t}$) and the grid power budget, defined as follows:
\begin{equation} \label{grid_budget}
\begin{aligned}
    P^{ID}_{G,t} = \frac{1}{\eta_{tr}} \Bigg( P^{\eta}_{B,t} + \Bigg[ 
    & \frac{P_{EV,t} - s^{-}_{t}}{\eta_{cp}} - \hat{P}^{\uparrow}_{PV,t} \cdot \eta_{pv}, \\ 
    & \frac{P_{EV,t} + s^{+}_{t}}{\eta_{cp}} - \hat{P}^{\downarrow}_{PV,t} \cdot \eta_{pv} 
    \Bigg] \Bigg)
\end{aligned}
\end{equation}

The optimization range is then divided in two time spans:
\begin{enumerate}
    \item Short-term time span, where a short-term PV forecaster is used to further refine the grid power budget with a smaller resolution ($\Delta j$) in $\frac{\Delta T}{\Delta j}$ range. This short term forecaster provides the median, a lower and an upper bound to the PV realization in the following $\frac{\Delta T}{\Delta j}$ time steps, defined as $\hat{P}_{PV,j}$, $\hat{P}^{\downarrow}_{PV,j}$ and $\hat{P}^{\uparrow}_{PV,j}$ respectively. Therefore, the grid power budget is upsampled substituting in Eq. \ref{grid_budget} $\hat{P}^{\downarrow}_{PV,t}$ and $\hat{P}^{\uparrow}_{PV,t}$ with $\hat{P}^{\downarrow}_{PV,j}$ and $\hat{P}^{\uparrow}_{PV,j}$, respectively, obtaining $P^{ID,up}_{G,j}$. The BESS power is upsampled obtaining $P^{\eta}_{B,j}$.

    On this range, also an a posteriori computation is performed to identify the cumulated maximum incentive over the short-term time span, as the additional savings/profit per unit of power:
    \begin{equation} \label{max_inc}
        D = max\left(a\cdot\mathcal{C}, \frac{\Delta R_{EV,T} - C_{BM,T}}{\sum^{t+\frac{\Delta T}{\Delta t}}_{t}P_{EV,t}\cdot \Delta t \cdot \frac{1}{\Delta T}}\right)
    \end{equation}
    Where the numerator is the additional revenues in the short-term from the EV charging ($\Delta R_{EV,T}$) and the BM transactions ($C_{BM,T}$):
    \begin{equation}
        \begin{aligned}
            & \Delta R_{EV,T} = \left(\sum^{t+\frac{\Delta T}{\Delta t}}_{t}P_{EV,t}\cdot \Delta t - P^{dp}_{EV,T}\cdot \Delta T \right) \cdot \mathcal{C}\\
            & C_{BM,T} = \bar{E}^+_{G,T} \cdot r_{short,T} + \bar{E}^-_{G,T} \cdot r_{long,T}
        \end{aligned}
    \end{equation}
    The denominator is the cumulated EV energy during the short-term span divided by the duration of this short term span $\Delta T$. The value for $D$ is the maximum between the computed value and a fraction of the current tariff that the CPO can decide ($a$).
    \item Long-term time span, where the BESS scheduling and the grid power budget remain with the $\Delta t$ time granularity up to the end of the refinement horizon $\mathcal{T}_{BM}$.
\end{enumerate}

\subsection{Real-time SG-ADMM} \label{core}

In real-time, let $N$ be the number of connected EVs at the beginning of the horizon $\mathcal{H}$. We defer those EVs arriving during the current horizon to the next horizon. The charging station and the $N$ EVs optimize two IOs, $f^*$ and $\hat{f}$ respectively. Both objective function are optimized through the same control variable, i.e. charging powers $\mathcal{P}$, but they have different optimal points, $\mathcal{P}^*$ for the CS and $\hat{\mathcal{P}}$ for the EVs. Due to these conflicting objective functions, the CS can be seen as the Leader and the EVs as the Followers in a Stackelberg Game framework, as previously introduced.

The CS needs to design in real-time an incentive mechanism such that each agent is willing to change its action to reach the Stackelberg equilibrium: let each i-th EV optimize a cost function $\Phi_{i}(\hat{f}(P_i),\theta_i)$, rather than $\hat{f}(P_i)$. By adjusting the parameter $\theta_i$, each EV is incentivized to reach its individual optimal point $P^*_i$ rather than $\hat{P}_i$. Hence, the incentive mechanism design problem can be formulated as a Stackelberg Game:

\begin{equation} \label{SG}
\begin{aligned}
&\hbox{Leader: } \Theta^* =\operatornamewithlimits{argmin}_{\Theta} \sum _{i=1}^{N}f^*_i(P_i)\\ 
&\hbox{N Followers: } P_i^*=\operatornamewithlimits{argmin}_{P_i} \Phi_{i}(\hat{f}(P_i),\theta^*_i) \\ 
&\hbox{Constraints: }\sum _{i=1}^{N} \frac{P_i}{\eta_{cp}} = C + s_L \text{ , } \sum_{i \in CC} P_i \leq P_{CC}\\ 
\end{aligned} 
\end{equation}
where $\Theta$ is the set of incentives and $\sum _{i=1}^{N} \frac{P_i}{\eta_{cp}} = {C} + s_L$ is a linear coupling constraint for both Leader's and Followers' games. In particular, at time $j$, $C$ is defined according to the ID refinement temporal upsampling for grid ($P^{ID,up}_{G,j}$) and BESS ($P^{\eta}_{B,j}$), and the PV power from the previous measurement ($P_{PV,j-1}$, considered in a persistent fashion) in this way:

\begin{equation} \label{constraint}
    C_j = P^{ID,up}_{G,j} \cdot \eta_{tr} - P^{\eta}_{B,j} + P_{PV, j-1} \cdot \eta_{pv}
\end{equation}
$s_L$ is instead a Leader defined slack, that will be discussed later. The second coupling constraint involve only sets of EV connected to the same charging column and imposes that the sum of the powers is less or equal to the CC rated power, for every CC. Since ADMM coupling constraints are introduced in the problem in the form of equality constraint, we can write that for all CC:
\begin{equation}
    \sum_{i \in CC}P_i + s_{CC} = P_{CC}
\end{equation}
Where $s_{CC}$ is the CC-specific (one per CC) slack variable, that will be discussed later.

The Leader controls the incentives for each Follower ($\theta_i$), while Followers control their own $P_i$. The idea is to design the cost function and the incentive parameters so that the Leader and the Follower can find the Stackelberg equilibrium. Starting from the cost function $\Phi_{i}(\hat{f}(P_i),\theta^k_i)$, it is the sum of a segmental Langrangian function for $P_i$ ($L_i(P_i,\lambda, \mu)$) and a purely individual part ($\phi^k_i(P_i)$), as in Eq. (\ref{incentive}).
\begin{equation} \label{inc.mech}
\begin{aligned}
    & \Phi_{i}(\hat{f}(P_i),\theta^k_i) = L_i(P_i,\lambda) + \phi^k_i(P_i)\\
     & =f^*_i(P_i)-\lambda^k \frac{P_i}{\eta_{cp}} - \mu_{CC}^k P_i+  \hat{f}_i(P_i) - \theta^k_i \cdot P_i \cdot \Delta j
     \end{aligned}
\end{equation}
where $\lambda^k$ and $\mu^k$ are the leader dual variables and the term $\theta^k_i \cdot P_i \cdot \Delta j$ incentivizes the EV to reach the Stackelberg equilibrium through a charging rate discount.

Before entering in the detail of the SG-ADMM algorithm, let's focus on the Leader's and Followers' objective functions:
\begin{itemize}
    \item In real-time the leader objective function ($\sum _{i=1}^{N}f^*_i(P_i)$) refer to a fair allocation of power ($\mathcal{P}$) among the followers. In particular, the goal is to reduce the absolute relative power deviation ($\alpha$ is a model hyperparameter).
    \begin{equation} \label{leader_specific}
        \sum _{i=1}^{N}f^*_i(P_i) = \alpha \cdot \sum_{i=1}^{N} \left( \frac{\vert P_{i} - P_{req,i}\vert}{P_{req,i}} \right)
    \end{equation}

\item Followers' objective ($\hat{f}_i(P_i)$) is to minimize the deviation from the requested power, that is the optimal trade-off between charging time and battery degradation computed and exposed by the BMS of the EV battery. This function is a piecewise function, where power lower than the required one reduce the predominant time-of-charge objective (weighted by the hyperparameter $\beta$) and power higher than the required one reduce the predominant battery degradation objective (weighted by the hyperparameter $\gamma$).
\begin{equation} \label{ev_specific}
 \hat{f}_i(P_i) = \begin{cases} \beta \cdot (P_{req,i} - P_i)^2 & \text{if } P_i \leq P_{req,i}\\
\gamma \cdot  \frac{SF_{P_i}}{SF_{P_{req,i}}} - 1 & \text{if } P_i > P_{req,i}
 \end{cases}
\end{equation}
It is defined so that the piecewise function holds continuity on the breakpoint $P_i=P_{req,i}$, that is also the minimum of the overall function with value of zero both from the right-hand and the left-hand limits.
\end{itemize}

We provide the two-layer nested iteration process of our SG-ADMM in Algorithm 1.

\begin{algorithm} 
\caption{SG-ADMM algorithm process}
\begin{algorithmic}[1]
    \State \textbf{Input:} $k=-1$, $s^0_L=0$, $P^{-1}_i=\hat{P}_i$
    \State \textbf{Output:} Optimal $P_i^*$, $\theta_i^*$, $\forall 1 \leq i \leq N$
    \While {\textbf{not} Outer Convergence}
        \State $k = k + 1$ and $t=0$
        \State (a) \textbf{Constraint update:}
        \State \hspace{1cm} $C^k = C + s^k_L$
        
        \While {\textbf{not} Inner Convergence}
        \State (1) Sequential follower update $\mathcal{P}^k(t+1)$
        \State (2) Inequality constraint update $s_{CC}^k(t+1)$
        \State (3) Leader duals update $\lambda^k(t+1), \mu_{CC}^k(t+1)$
        \State (4) $t = t + 1$
        \State (5) Check Inner Convergence criterion
        \State (6) Penalty parameter update $\rho(t+1)$
    \EndWhile
    \State (b) \textbf{Leader's Incentive and Slack Design:}
    \State \hspace{1cm} $\Theta^{k+1}, s^{k+1}_L$ as in Figure \ref{outer}
    \State (c) \textbf{Check Outer Convergence criterion}
    \EndWhile
    \State $k = k-1$
    \State \textbf{Result:} $P_i^* = P_i^k$, and $\theta_i^* = \theta_i^k$, $\forall 1 \leq i \leq N$
\end{algorithmic}
\end{algorithm}

It is based on two nested loops, where the outer is the Stackelberg Game that initiates from the inner loop ADMM optimization. The algorithm enters the outer loop provided that the outer convergence criterion is not satisfied. Once in the outer loop, the constraint is updated as per line 5-6. Similarly, the algorithm enters the inner loop provided that the inner convergence criterion is not satisfied. The inner loop consists of four steps, i.e. the sequential follower update, the leader dual update, the inner loop iteration update, the check of the inner convergence criterion and the penalty parameter update. Thus, at each step k, given leader’s strategy ($\Theta^k$ and $C^k$), the follower optimizes $\Phi_i(\hat{f}_i(P_i),\theta^k_i)$ to solve the Follower game in Eq. \ref{SG}, through ADMM. Once the inner convergence criterion is satisfied, the leader incentive and slack design is performed to solve the Leader game in Eq. \ref{SG} (and it will be described later) and the outer convergence criterion is checked.

\begin{figure*}[h]
    \centering
    \includegraphics[width=0.86\linewidth]{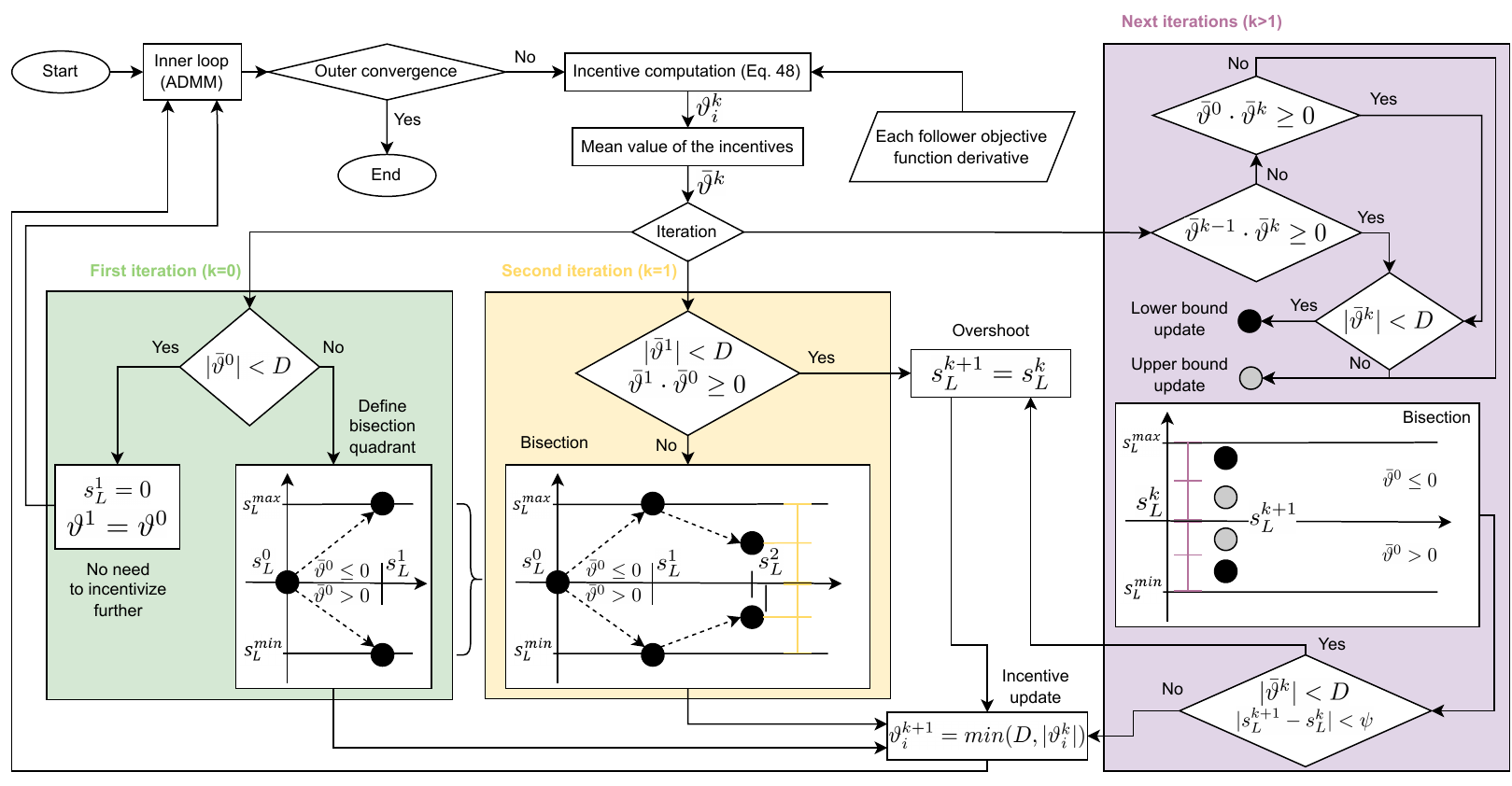}
    \caption{Outer loop based on the bisection method: the first iteration (green) either defines the quadrant or keep zero the slack, the second iteration (yellow) either bisects or - in case of overshooting - reestablish the previous slack, the following iterations (purple) help the outer loop to converge to the lower value of slack with a feasible incentive.}
    \label{outer}
\end{figure*}

Starting from the inner loop, here the main steps:
\begin{enumerate}
    \item Sequentially follower's update:
    \begin{equation} \label{sfu}
    \begin{aligned} 
        & P_i^k(t+1) = \operatornamewithlimits{argmin}_{P_i} \Phi_{i}(\hat{f}(P_i),\theta^k_i)\\ & +\frac{\rho}{2} \left\| \frac{1}{\eta_{cp}}\left(\sum_{j=1}^{i-1} P_j^k(t+1) + P_i + \sum_{j=i+1}^{N} P_j^k(t)\right) - C^k\right\|_2^2\\
        & +\frac{\rho}{2} \left\| \left(P_i + \sum_{j \in CC} P_j^k(t +  \mathbf{1}_{(j < i)}) + s^k_{CC}(t)\right) - P_{CC}\right\|_2^2\\
    \end{aligned}
    \end{equation}
    where $\mathbf{1}_{(j < i)}$ is an indicator function that identifies if the $j^{\text{th}}$ EV update already occurred or not. $\rho$ is the augmented Lagrangian penalty parameter, that is updated at the end of the loop according to \cite{Boyd_ADMM} (Eq. 3.13) for improving convergence (written as $\rho$ - and not $\rho(t+1)$ for brevity).
    \item Inequality constraint update:
    \begin{equation}
    \begin{aligned}
        & s_{CC}^k(t+1) = max\left(0, P_{CC} - \sum_{i \in CC}P^k_i(t+1)\right)\\
        & P^k_{CC}(t+1) = \sum_{i \in CC}P^k_i(t+1) + s_{CC}^k(t+1)
    \end{aligned}
    \end{equation}
    Where $P^k_{CC}(t+1)$ is an auxiliary variable introduced for brevity.
    \item Leader's duals update:
    \begin{equation} \label{ldu}
    \begin{aligned}
        & \lambda^k(t+1) = \lambda^k(t) - \rho \left( \sum_{i=1}^{N} \frac{P_i^k(t+1)}{\eta_{cp}}  - C^k \right)\\
        & \mu_{CC}^k(t+1) =  \mu_{CC}^k(t)- \rho \left(P^k_{CC}(t+1) - P_{CC} \right)
    \end{aligned}
    \end{equation}
    \item Inner Loop Convergence Criterion:
    \begin{equation} \label{icc}
    \begin{aligned}
    & \|\mathbf{r}^k(t+1)\|_2 \leq \epsilon_{primal} \text{,} \;\|\mathbf{s}^k(t+1)\|_2 \leq \epsilon_{dual}\\
    &\mathbf{r}^k (t+1) =
\begin{bmatrix}
\sum_{i=1}^N \frac{P_i^k (t+1)}{\eta_{cp}} - C^k \\[8pt]
 P^k_{CC}(t+1) - P_{CC}
\end{bmatrix}\\
& \mathbf{s}^k (t+1) = \rho
\begin{bmatrix}
\sum_{i=1}^{N} \frac{1}{\eta_{cp}} \left( P_i^k(t+1) - P_i^k(t) \right) \\[8pt]
 s^k_{CC}(t+1)-s^k_{CC}(t)
\end{bmatrix} \\
\end{aligned}
\end{equation}
where $\mathbf{r}^k$ and $\mathbf{s}^k$ are the primal and dual residuals' vectors, respectively.
$\epsilon_{primal}$ and $\epsilon_{dual}$ are the primal and dual residual tolerance. They are calculated as follows, following indications from \cite{Boyd_ADMM}:
\begin{equation}
\begin{aligned}
    &\epsilon_{primal} = \epsilon_{abs} + \epsilon_{rel} \cdot max \left(C^k, \sum_{i=1}^N \frac{P_i^k (t+1)}{\eta_{cp}}\right)\\
    &\epsilon_{dual} = \epsilon_{abs} + \epsilon_{rel} \cdot \lambda^k(t+1)\\ 
\end{aligned}
\end{equation}

\item Penalty parameter update:
\begin{equation}
    \rho(t+1) = \begin{cases}
        \tau_{\rho} \rho(t) \quad\text{if } \|\mathbf{r}^k(t+1)\|_2 >\mu \|\mathbf{s}^k(t+1)\|_2\\
         \frac{\rho(t)}{\tau_{\rho}} \quad\text{if } \|\mathbf{s}^k(t+1)\|_2 >\mu \|\mathbf{r}^k(t+1)\|_2\\
         \rho(t) \quad \text{otherwise}
    \end{cases}
\end{equation}
Where $\tau_{\rho}$ and $\mu$ are ADMM hyperparameters, fixed to 2 and 10 respectively as in \cite{Boyd_ADMM}.
\end{enumerate}

The outer loop comprises the followers feedback on the control variables $\mathcal{P}$ and the corresponding update of the Leader incentives and slack variable. The incentive update is set by the leader to the current marginal cost of each follower multiplied by a model hyperparameter $\delta$:
\begin{equation}
    \theta^{k+1}_i = \delta \cdot \nabla_{P_i} \hat{f}_i(P_i)
\end{equation}

The Leader slack update is performed on the principle of the bisection method on the feasible slacks, where we search for the lower slack – in absolute value – that corresponds to an incentive within the acceptable range. We show this principle in Figure \ref{outer}, where the bisection method uses the historical and the current incentives to update the leader slack until outer convergence. $D$ is the maximum incentive defined by the Leader (Eq. \ref{max_inc}). The slack variable is included in $[0; s_L^{max}]$ or in $[s_L^{min}; 0]$, whether the initial average value $\bar{\theta}^1$ is negative or positive, respectively. The maximum and the minimum slack are defined as follows:
\begin{equation} \label{max_slacks}
    \begin{aligned} 
        & s_L^{max} = \frac{s^{+}_j}{\eta_{cp}} + (P_{PV, j-1} - \hat{P}^{\downarrow}_{PV})\cdot \eta_{pv}\\
        & s_L^{min} = -\frac{s^{-}_j}{\eta_{cp}} + (P_{PV, j-1} - \hat{P}^{\uparrow}_{PV})\cdot \eta_{pv}
    \end{aligned}
\end{equation}

The outer loop convergence criterion is the following:
\begin{equation}
\begin{aligned}
    & ||L(\mathcal{P}^{k+1}, \lambda^{k+1}) - L(\mathcal{P}^{k}, \lambda^{k})|| \leq \varepsilon\\
    & L(\mathcal{P}^k, \lambda^k) = \sum^{N}_{i=1} L_i(P^k_i,\lambda^k) + \lambda^k C^k \\
    &+ \sum_{CC=1}^{N_{CC}} \mu^k_{CC} s_{CC}^k + \frac{\rho}{2} \left\| \sum_{i=1}^{N} \frac{P^k_i}{\eta_{cp}}  - C^k\right\|_2^2\\
    & + \frac{\rho}{2}\sum_{CC=1}^{N_{CC}} \left\| \sum_{i \in CC} P^k_i + s^k_{CC}  - P_{CC}\right\|_2^2
\end{aligned}
\end{equation}

\section{Conclusions} \label{concl_part1}
Since decentralized optimization might suffer from the lack of incentives that steer the agents' IOs towards the central controller optimum, in this work we propose a novel application of the SG-ADMM algorithm originally proposed in \cite{CORE}, applied to the real-time control of an EVCS. This work provides a literature review on the use of SG-ADMM for single-leader multi-followers non-cooperative games and formulates it inside a hierachical multi-layered EMS. In the first part of this two-part paper, we draw up the overall EMS formulation focusing on the modifications to SG-ADMM. Indeed, the original algorithm has been tweaked to accomodate the problem formulation and improve the overall convergence. The inner loop, based on ADMM, sets the followers' demand in response to the leader's incentives. The outer loop, a Stackelberg game, consists in the Leader incentive and constraint update, performed by means of a bisection method trading off the coupling constraint, i.e. the available power, and incentive provision.

\bibliographystyle{IEEEtran}
\bibliography{literature}

\end{document}